\documentclass[useAMS,usenatbib]{mn2e}
\usepackage{graphicx} \graphicspath{{Figs/}}
\usepackage{txfonts}
\usepackage{Journals}
\pdfoutput=1
\def\eq#1{\begin{equation} #1 \end{equation}}

\def\E#1{\hbox{$10^{#1}$}}
\def\sub#1{_{\rm #1}}
\def\about {\hbox{$\sim$}}
\def\deg   {\hbox{$^\circ$}}
\def\x     {\hbox{$\times$}}
\def\xc    {\hbox{$x_{\rm c}$}}
\def\xw    {\hbox{$x_{\rm w}$}}

\def\Ri    {\hbox{$R_{\rm in}$}}
\def\Ro    {\hbox{$R_{\rm out}$}}
\def\vi    {\hbox{$\v_{\rm in}$}}
\def\vo    {\hbox{$\v\sub{out}$}}
\def\vp    {\hbox{$\v\sub{p}$}}
\def\vrot   {\hbox{$\v\sub{rot}$}}
\def\tin    {\hbox{$\theta\sub{in}$}}
\def\tout   {\hbox{$\theta\sub{out}$}}
\def\T     {\hbox{$\tau_{\rm T}$}}
\def\Tx    {\hbox{$T_{\rm x}$}}
\def\nc    {\hbox{$n_{\rm crit}$}}
\def\Dnu   {\hbox{$\Delta\nu_{\rm D}$}}
\def\Dv    {\hbox{$\Delta\v_{\rm D}$}}

\def\Sbar  {\hbox{$\bar S$}}
\def\Imax   {\hbox{$I\sub{max}$}}
\def\Imin   {\hbox{$I\sub{min}$}}
\def\lmax   {\hbox{$\ell\sub{max}$}}
\def\lmin   {\hbox{$\ell\sub{min}$}}

\let\DS=\displaystyle

\font\math = cmmi10
\def\m#1{\hbox{\math \char'#1}} 
\def\v{\m{166}}

\title[Non-Kinematic Double Peaks]
{Rotating Disks and Non-Kinematic Double Peaks}

\author[Elitzur, Asensio Ramos \& Ceccarelli]
       {Moshe Elitzur$^1$,
        Andr\'es Asensio Ramos$^2$ and
        Cecilia Ceccarelli$^3$\\
 $^1$Department of Physics \& Astronomy, University of Kentucky,
    Lexington, KY 40506, USA; moshe@pa.uky.edu \\
 $^2$Instituto de Astrof\'{\i}sica de Canarias, 38205, La Laguna, Tenerife,
    Spain; aasensio@iac.es \\
 $^3$Laboratoire d'Astrophysique de l'Observatoire de Grenoble,
    BP 53, 38041 Grenoble, Cedex 9, France; Cecilia.Ceccarelli@obs.ujf-grenoble.fr
}

\date{Accepted 2012 February 7.  Received 2012 January 13; in original form 2011 December 6}
\pagerange{\pageref{firstpage}--\pageref{lastpage}} \pubyear{2012}

\begin{document}
\label{firstpage}

\maketitle

\begin{abstract}
Double-peaked line profiles are commonly considered a hallmark of rotating
disks, with the distance between the peaks a measure of the rotation velocity.
However, double-peaks can arise also from radiative transfer effects in
optically thick non-rotating sources. Utilizing exact solutions of the line
transfer problem we present a detailed study of line emission from
geometrically thin Keplerian disks. We derive the conditions for emergence of
kinematic double peaks in optically thin and thick disks, and find that it is
generally impossible to disentangle the effects of kinematics and line opacity
in observed double-peaked profiles. Unless supplemented by additional
information, a double-peaked profile alone is not a reliable indicator of a
rotating disk.  In certain circumstances, triple and quadruple profiles might
be better indicators of rotation in optically thick disks.

\end{abstract}


\section{Introduction}
\label{sec:introduction}

Rotating disks are key elements in numerous astrophysical systems.
Protoplanetary disks around young stellar objects (YSOs) are where planets
eventually form, and thus are of paramount importance in all theories of planet
formation. Mass transfer in cataclysmic variables (CVs) proceeds through an
accretion disk around the white dwarf. In active galactic nuclei (AGNs), the
accretion disks are the conduit of material to the central black hole and the
source of continuum emission at all wavelengths shorter than IR. Unfortunately,
small angular sizes preclude direct imaging of the disks in most CVs and YSOs,
and virtually all AGNs; only a handful of AGNs provide direct evidence for
Keplerian disks through interferometric observations of water masers
\citep[see][]{Greenhill07}. In the absence of imaging, a double-peaked line
profile is often taken as evidence for an inclined rotating disk, with the
distance between the peaks a measure of the rotation velocity. For example, the
detection of double-peaked profiles has been used to establish the presence of
disks in AGNs \citep[see][and references therein]{Eracleous09}, and the
analysis of double-peaked profiles has been used to deduce the mass of the
central object from the derived Keplerian rotation (\cite{Zhang08}).

In the widespread reliance on the double-peaked profile as a marker of rotating
disks it has been overlooked that such a profile may actually result not from
disk rotation but, instead, from purely radiative transfer effects in optically
thick, non-rotating systems. As has long been known, the simplest line transfer
problem involving a uniform, quiescent slab without any large-scale motions
leads to double-peaked profiles at large optical depths \citep[see][and
references therein]{Avrett65}. Unfortunately, the existence of these {\em
non-kinematic double peaks} has been largely overlooked.  Instead, the
\cite{Horne86} seminal work introduced flat-top line profiles for the optically
thick segments of rotating disks, an approximation that ignores the spatial
variation of the source function. A number of recent detailed calculations of
line emission from protoplanetary systems did produce double-peaked profiles in
some optically thick face-on disks \citep{Pavlyuchenkov07, Cernicharo09,
Ceccarelli10}, but the implications of these results were not discussed. The
significance of non-kinematic double peaks has been noted only relatively
recently by \cite{Hummel00} in the context of B[e] stars.
%
%

Our goal here is to present a general study of line emission from rotating
disks to determine whether and when the effects of kinematics and line opacity
can be disentangled in observed double-peaked profiles. Following
\citeauthor{Horne86} we consider a geometrically-thin Keplerian disk and
compute its line emission profile for a large range of optical depths with
detailed radiative transfer calculations. The radiative transfer problem for a
non-rotating geometrically thin uniform disk is identical to that for an
infinite uniform slab when neglecting radiation propagation in the plane. In
\S\ref{sec:Slab} we demonstrate the emergence of non-kinematic double peak
profiles, without any large scale motions, in the line emission from optically
thick uniform slabs; in Appendix \ref{sec_ap:slab} we derive some simple
analytic expressions that approximate adequately the exact calculations. In
\S\ref{sec:Disk} we present the solution for Keplerian disks, which shows
double peak line emission from face-on viewing. We study the properties of
kinematic double peaks, which are introduced by the disk rotation, and compare
them with the non-kinematic ones. Section \S\ref{sec:discussion} contains a
discussion.

\section{Non-Kinematic Double Peaks}
\label{sec:Slab}

To demonstrate the formation of double-peaked profiles in the absence of any
large scale motions we start with the simplest problem involving the smallest
possible number of free parameters---a uniform slab with constant temperature
and density and no radiation other than line radiation generated internally by
collisions. We solve the problem with the Coupled Escape Probability (CEP);
this is an exact method we developed recently for the line transfer problem
that offers great speed advantages over traditional computational approaches
\citep{CEP06}. We consider line emission from a transition between upper level
$u$ and lower level $l$ separated by energy $E_{ul} = h\nu_0$. The transition
properties are fully specified for line transfer calculations by the parameter
$\epsilon$, the photon destruction probability defined through
\eq{\label{eq:epsilon}
    {\epsilon\over 1 - \epsilon}
      = {C_{ul}\over A_{ul}}\left(1 - e^{-E_{ul}\!/kT}\right)
      \equiv {n\over\nc}
}
Here $C_{ul}$ and $A_{ul}$ are, respectively, the collisional de-excitation
rate and the spontaneous emission coefficient, $T$ is the gas temperature, $n$
the density and \nc\ the transition critical density corrected with the
Boltzmann factor \citetext{see, e.g., eq.\ 24 in \citealt{CEP06}}. Vertical
position in the slab can be specified by the coordinate $\tau$, the optical
depth at line center which varies from 0 to \T\ between the two faces. Denote
by \Dnu\ the frequency width of the thermal motions then the optical depth at
every frequency is
\eq{\label{eq:tau}
    \tau(x) = \tau\Phi(x), \qquad
    \hbox{where}\quad x = {\nu - \nu_0\over\Dnu}
}
and where $\Phi(x)$ is a profile normalized to unity at line center. The
problem is fully specified by just two free parameters, $\epsilon$ and \T, and
the line absorption profile. In the numerical calculations here we take $\Phi$
as Gaussian, $\exp(-x^2)$, but the results are applicable to other profiles,
for example Lorentzian shape.

\begin{figure}
  \centering
  \includegraphics[width=\hsize]{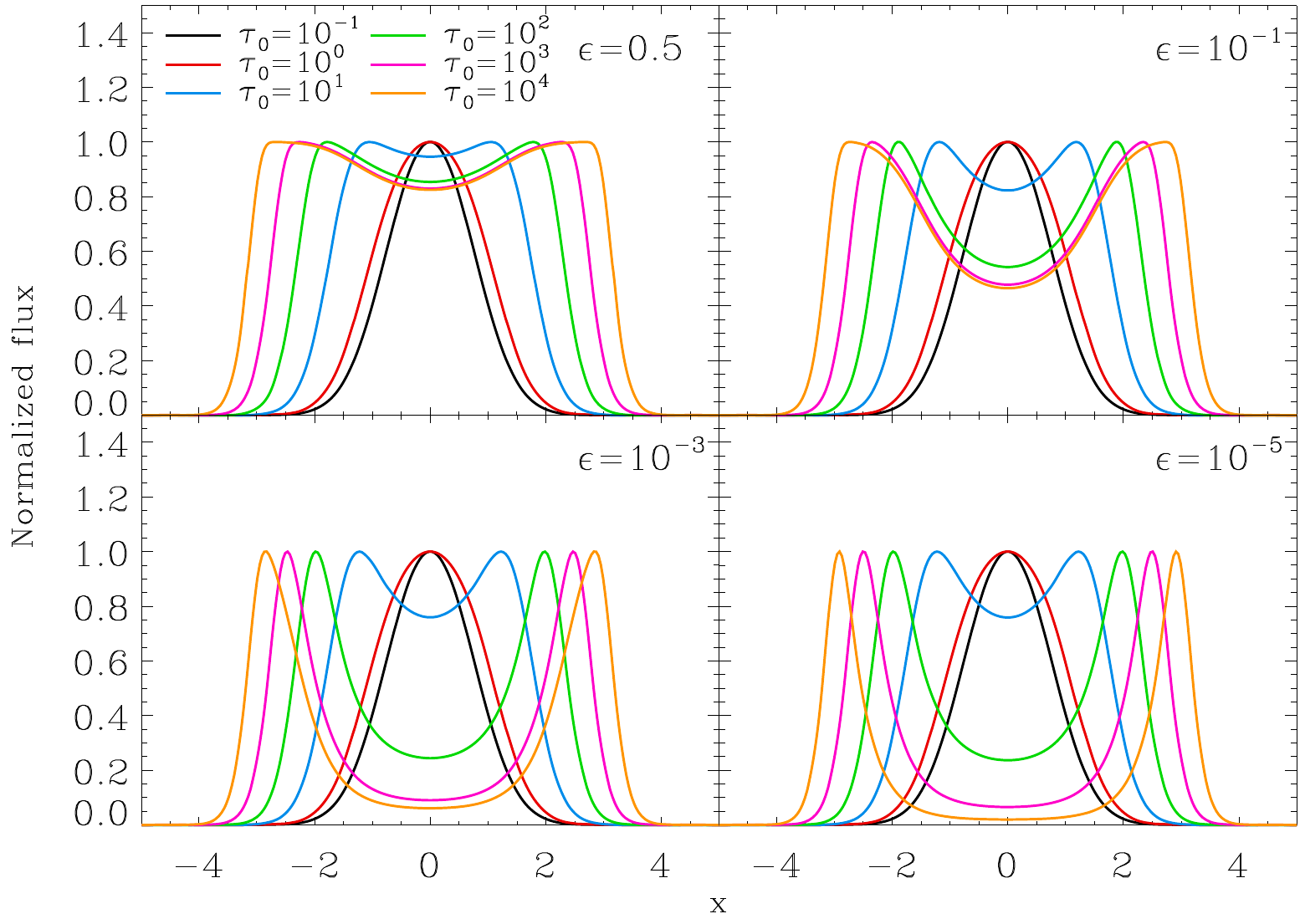}

\caption{Frequency profiles $\Psi(x)$ of line radiation from a uniform slab,
normalized to a unity peak flux, from exact numerical calculations with the CEP
method. Each panel corresponds to a different value of the photon destruction
probability $\epsilon$ (see eq.\ \ref{eq:epsilon}). Different lines correspond
to different values of \T, the line-center slab optical depth, as marked.
}
 \label{fig:slab}
\end{figure}

\begin{figure}
  \centering
  \includegraphics[width=\hsize]{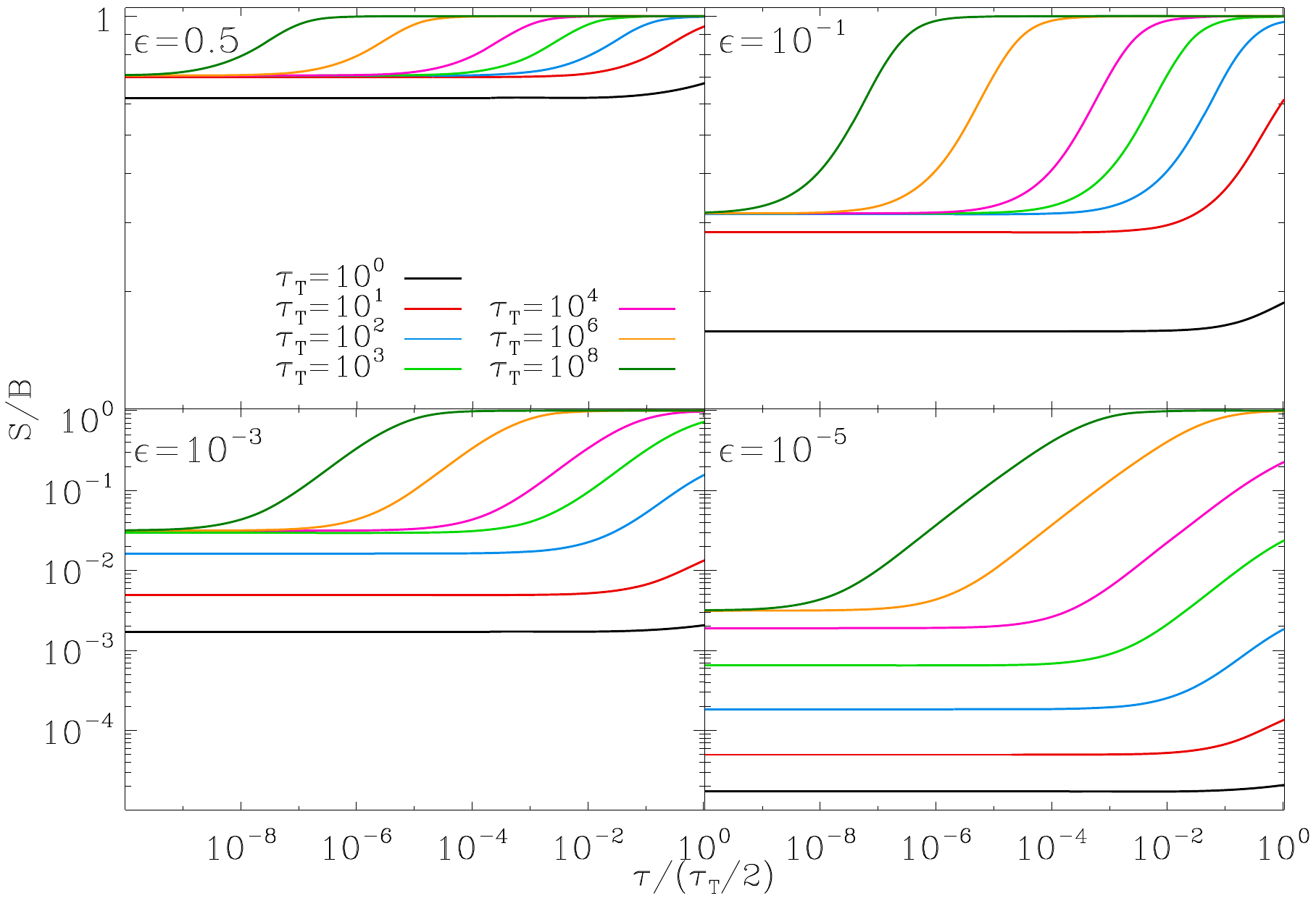}

\caption{Variation of the source function, normalized by the Planck function of
the slab temperature, with optical depth into the slab for the models shown in
figure \ref{fig:slab}. The profiles are symmetric about the slab midplane,
which is the endpoint of the horizontal axis in each panel. Because the axis is
scaled with the overall optical depth, the same point in each panel indicates a
different distance from the surface. Note also that $S/B = 1$ implies a
thermalized transition ($\Tx = T$).
}
 \label{fig:Tx_Slab}
\end{figure}

Figure \ref{fig:slab} shows the frequency profiles $\Psi(x)$ of the line
radiation leaving the slab vertically for different sets of the free parameters
$\epsilon$ and \T. In optically thin slabs, $\Psi(x) = \Phi(x)$.  Double-peaked
profiles emerge when $\T \ge 10$ and become more pronounced as $\epsilon$ gets
smaller, i.e., as the density decreases away from the critical density. As
noted in the Introduction, double-peaked profiles in optically thick sources
are a well known radiative transfer effect \citep[e.g.,][]{Avrett65}. The
explanation is rather simple. Consider a slab whose line center optical depth
to the midplane exceeds unity. Then the frequency \xc\ where
$\frac12\T\Phi(\xc) = 1$ defines the line core, the spectral region $|x| < \xc$
in which each half of the slab is optically thick. Similarly, the frequency
\xw\ where $\T\Phi(\xw) = 1$ defines the line wings---the entire slab is
optically thin at $|x| > \xw$. For the Doppler profile, the explicit
expressions for these two frequencies are
\eq{\label{eq:x1}
    \xw = \sqrt{\ln\T}\,, \qquad
    x_{\rm c}^2 = x_{\rm w}^2 - \ln 2
}
At the line core, $|x| < \xc$, each half of the slab is optically thick and the
radiation emerges from an optical distance of only \about\ 1 from each surface.
Thus the line core intensity is roughly the value of the source function $S$ at
position $\tau = e^{x^2}$ (see eq.\ \ref{eq:I(x)}); that is, the core spectral
shape reflects the spatial variation of $S$, or, equivalently, the line
excitation temperature \Tx\ since $S = B(\Tx)$ where $B$ is the Planck function
\citep[e.g.,][]{MasersBook}. Because of photon trapping, \Tx\ is maximal on the
slab midplane, declining toward each surface. So the core intensity decreases
from a peak at $|x| = \xc$, corresponding to the magnitude of \Tx\ on the
midplane, to a minimum at line center, where it reflects the excitation
temperature at distance $\tau = 1$ from the surface. At the line wings, $|x| >
\xw$, the slab is optically thin everywhere and its emission is simply $I(x) =
\Sbar\T\Phi(x)$, where \Sbar\ is the value of the source function at some
intermediate point inside the slab (see appendix \ref{sec_ap:slab}); that is,
the spectral shape of emission at the wings follows the line profile, $\Psi(x)
= \Phi(x)$. The intensity now decreases as $x$ increases because the entire
slab is fully visible and the optical depth decreases away from line center.

Figure \ref{fig:Tx_Slab} shows the spatial profile of $S/B$, where $B(T)$ is
the Planck function of the (constant) slab temperature, vs optical depth into
the slab for a large range of $\epsilon$ and \T. This spatial variation is
directly reflected in the properties of the emerging radiation profile (fig.\
\ref{fig:slab}). Once the level populations reach thermal equilibrium, $\Tx =
T$ and $S/B = 1$. When $\T \gg 1/\epsilon$, the transition is thermalized at
slab center as is evident from the figure (see also eq.\ \ref{eq:S_fits}).
Approaching the slab surface, photon escape reduces \Tx\ and the source
function is decreasing, producing the intensity dip at line center. When \T\
increases, thermalization spreads from the slab midplane toward its surfaces,
larger portions of the source have $\Tx \simeq T$ and the peaks flatten out.
This flattening becomes more pronounced as $\epsilon$ increases, i.e., $n$
approaches \nc. When $\T < 1/\epsilon$, the level populations do not reach
thermal equilibrium even at slab center, but photon trapping ensures that \Tx,
and the source function, is always higher there than at the surface.

\begin{figure}
  \centering
  \includegraphics[width=\hsize]{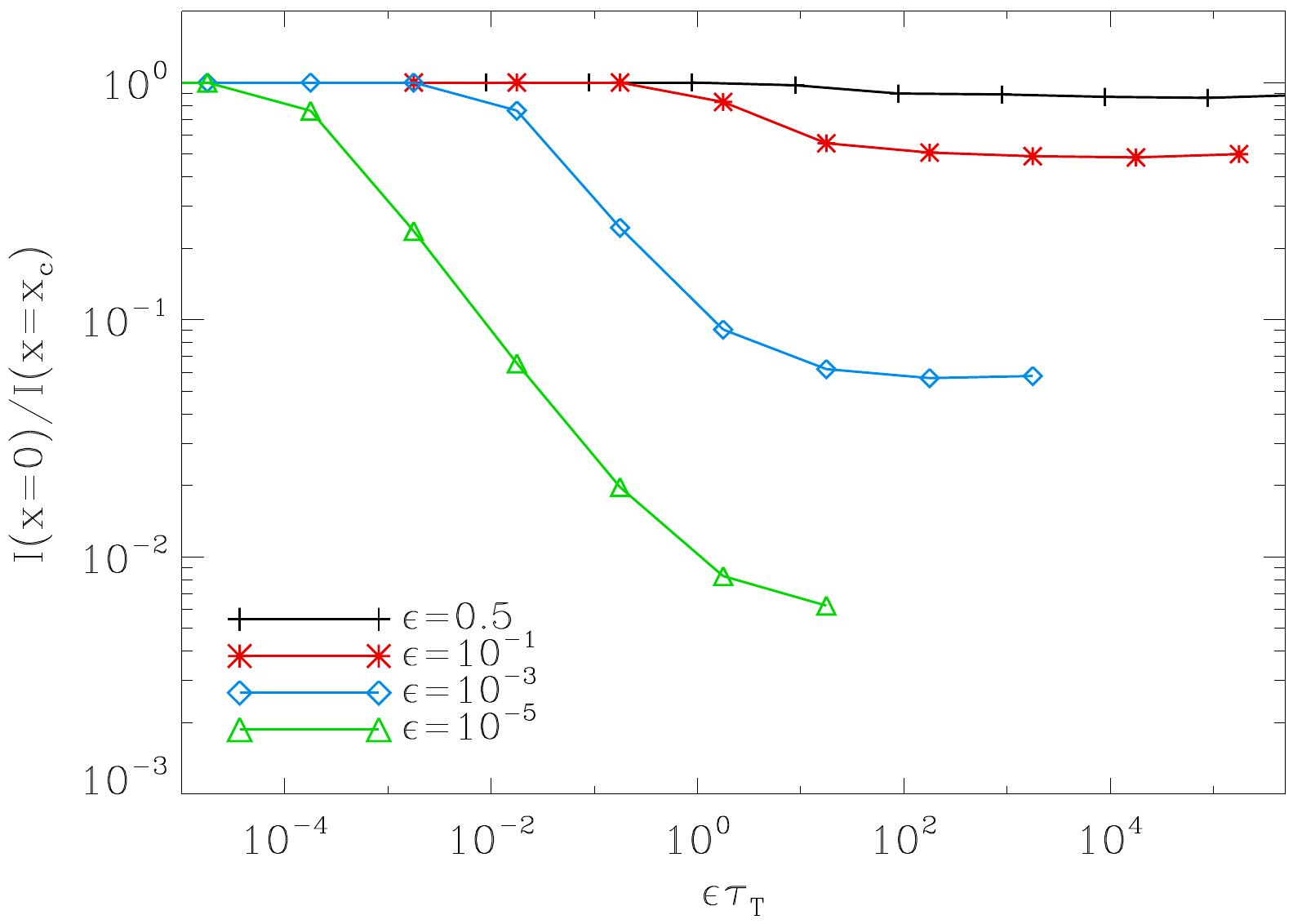}
\caption{Variation of the depth of the line profile central dip with slab
optical thickness for various values of $\epsilon$; the horizontal axis
reflects the problem's scaling properties. When $\T \gg 1/\epsilon$, the depth
of the dip saturates at a value that varies roughly as $\sqrt{\epsilon}$ (see
eq.\ \ref{eq:contrast0}).
}
 \label{fig:Dip}
\end{figure}

The behavior of the central dip seen in fig.\ \ref{fig:slab} is readily
explained by the spatial variation of the source function. The depth of the dip
can be characterized by the ratio of intensities at line center and at the
peaks. Figure \ref{fig:Dip} shows the variation of this ratio with optical
thickness for various values of $\epsilon$; because of the scaling properties
of the problem (see eq.\ \ref{eq:S_fits}), the horizontal axis is labeled with
$\epsilon\T$. For every $\epsilon$, the dip reaches maximal depth when $\T \gg
1/\epsilon$. The dip essentially disappears at $\epsilon = 0.5$, corresponding
to $n = \nc$, because the source function thermalizes throughout most of the
slab; in this regime, $S(\tau)$ is nearly flat across the line core (see fig.\
\ref{fig:Tx_Slab}). In Appendix \ref{sec_ap:slab} we discuss these issues
further and derive semi-analytic approximations that describe adequately the
exact numerical results; in particular, fig.\ \ref{fig:profiles} reproduces the
line profiles quite accurately with simple analytic expressions. The depth of
the central dip is accurately described by eq.\ \ref{eq:dip}; to a good degree
of approximation, its large-\T\ limit is
\eq{\label{eq:contrast0}
    {\Imin\over\Imax} = {I(x = 0)\over I(x = \xc)}
        \mathrel{{\mathop\simeq_{\DS \epsilon\T \gg 1}}} \sqrt{\epsilon}
}
(see eq.\ \ref{eq:dip2}). This approximation describes reasonably well the
results shown in figure \ref{fig:Dip}.

This standard problem demonstrates that double-peaked profiles need not require
any large scale motion because they are generated naturally by radiative
transfer effects. We will refer to these as {\em non-kinematic double peaks}.
In very optically thick sources, the emergent radiation peaks at the frequency
shifts $x \simeq \pm \xc$, producing a peak separation of
\eq{\label{eq:peaks}
    \Delta\nu_{\rm peak} \simeq 2\xc\Dnu \simeq 2\Dnu\sqrt{\ln\T}
}
Outside the peaks, the intensity falloff traces the shape of the Doppler wings.
Inside the peaks, the intensity falloff toward line center reflects the spatial
decline of the source function, i.e., the line excitation temperature, when the
surface is approached.

\section{Rotating Disk Line Profiles}
\label{sec:Disk}

\subsection{Basic Theory}
\label{sec:Basic}

Consider a geometrically-thin disk, extending from an inner radius \Ri\ to $\Ro
= Y\Ri$. The intensity toward angle $i$ from the disk normal is $I_\nu(R,i)$,
where $R$ is axial radius. Then the flux observed at distance $D$ and viewing
angle $i\ (< 90\deg)$ is $L_\nu(i)\x(\cos i/D^2)$, where
\eq{\label{eq:first}
   L_\nu(i) = \int I_\nu(R,i)\,dA
}
is the disk monochromatic luminosity in direction $i$. The observed flux is
obtained from an integration over the circular surface area even though the
disk image is an ellipse; the reason is that projection on the plane of the sky
distorts shapes but preserves relative sizes of area elements because all are
scaled by the common factor $\cos i$.

\begin{figure}
  \centering
  \includegraphics[width=0.7\hsize,clip=]{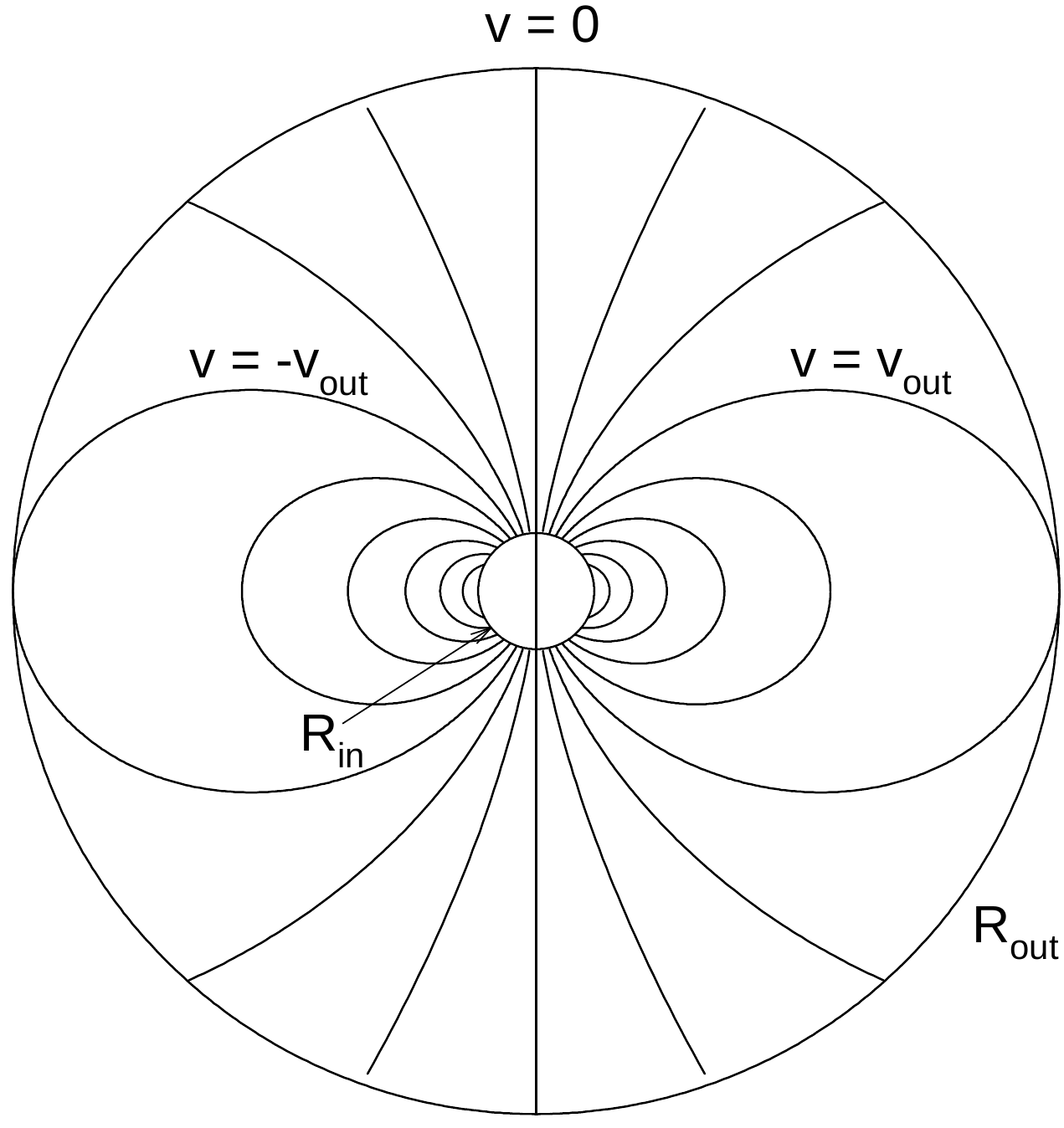}
\caption{Contours of constant line-of-sight (los) velocity \v\ for a Keplerian
disk between radii \Ri\ and \Ro. The observer is along the \v\ = 0 contour. In
the 1st quadrant, the los velocity increases along successive contours
clockwise from the top.
}
 \label{fig:Contours}
\end{figure}

The line intensity of a non-rotating disk is $I_\nu(R,i) = I_0(R,i)\Psi(\nu -
\nu_0,R,i)$, where $I_0$ is the brightness at line center and $\Psi$ is the
spectral shape of the emerging line emission. When the line is optically thin,
$\Psi(\nu) = \Phi(\nu)$. Determining $\Psi$ for optically thick lines requires
a full radiative transfer calculation; a sample of the results of such
calculations is shown in fig.\ \ref{fig:slab}. In general, $\Psi$ varies with
$i$ because of the variation of optical depth with inclination; it can also
vary with $R$ if the vertical optical depth varies with radius. Switching from
frequency to equivalent velocity $\v = c(\nu_0 - \nu)/\nu_0$, the profile
becomes $\Psi(\v,i)$, centered on $\v = 0$. Now set the disk in Keplerian
rotation $\vrot(R)$. For viewing angle $i$, the line-of-sight (los) component
of the local rotation velocity of a point at radius $R$ and angle $\theta$ from
the midline is
\eq{\label{eq:vrot}
   \v_z(R,\theta,i) = \vo\left(\Ro\over R\right)^{1/2}\cos\theta,
   \quad \hbox{where} \quad
   \vo = \vrot(\Ro)\sin i
}
The line emission profile is now centered on $\v_z(R,\theta,i)$ and the
monochromatic luminosity of a Keplerian disk is
\eq{\label{eq:basic}
   L(\v,i) = \int I_0(R,i)\,\Psi\!\left(\v - \v_z[R,\theta,i]\right)\,R\,dR\,d\theta
}
This is the basic expression for the spectral shape of disk line emission. In
the case of a uniform disk the emerging profile involves simply the areal
integration of the intrinsic emission profile $\Psi$ with the position
variation introduced by the Keplerian rotation. We now present calculations for
uniform disks ($R$-independent $I_0$ and $\Psi$) and various forms of $\Psi$.
Although one can expect flaring and non-uniformity in physical disks, the
uniform thin disk can be expected to faithfully reproduce the essence of the
radiative transfer effects common to all disks.

\subsection{Kinematic Double Peaks---Optically Thin Disks}
\label{sec:thin}

The locus of all points whose emission is centered on the same frequency \v\ is
the constant los-velocity curve  $\v_z = \v$ on the surface of the disk. From
eq.\ \ref{eq:vrot}, for $\v \ne 0$ this curve obeys
\eq{\label{eq:contours}
   R = \Ro\left(\vo\over \v\right)^2\cos^2\theta
}
Figure \ref{fig:Contours} shows the contour plots of these curves for some
representative velocities. In the case of a rectangular emission profile
$\Psi$, the emission at \v\ is proportional to the area of a strip centered on
the corresponding contour. When the rectangular $\Psi$ is so narrow that it can
be approximated with a $\delta$-function, the emission becomes roughly
proportional to the contour length. The shortest contour is the straight line
at \v\ = 0. With \v\ increasing, as long as $\v < \vo$ the contours remain
open, their curvature increases and with it their length. The first fully
closed contour is the one for \vo, corresponding to the smallest los velocity
on the disk midline. Increasing \v\ further corresponds to smaller radial
distances on the midline, and the contours are now shrinking, finally reaching
the single point at the midline intersection with the inner radius where $\v =
\vi\ (= \vo\sqrt{Y})$. The contour lengths are calculated in closed form in
Appendix \ref{sec_ap:disk} and plotted in figure \ref{fig:lengths} for various
values of $Y$ using the analytic result in eq.\ \ref{eq:ell}.

\begin{figure}
\centering
  \includegraphics[width=\hsize,clip=]{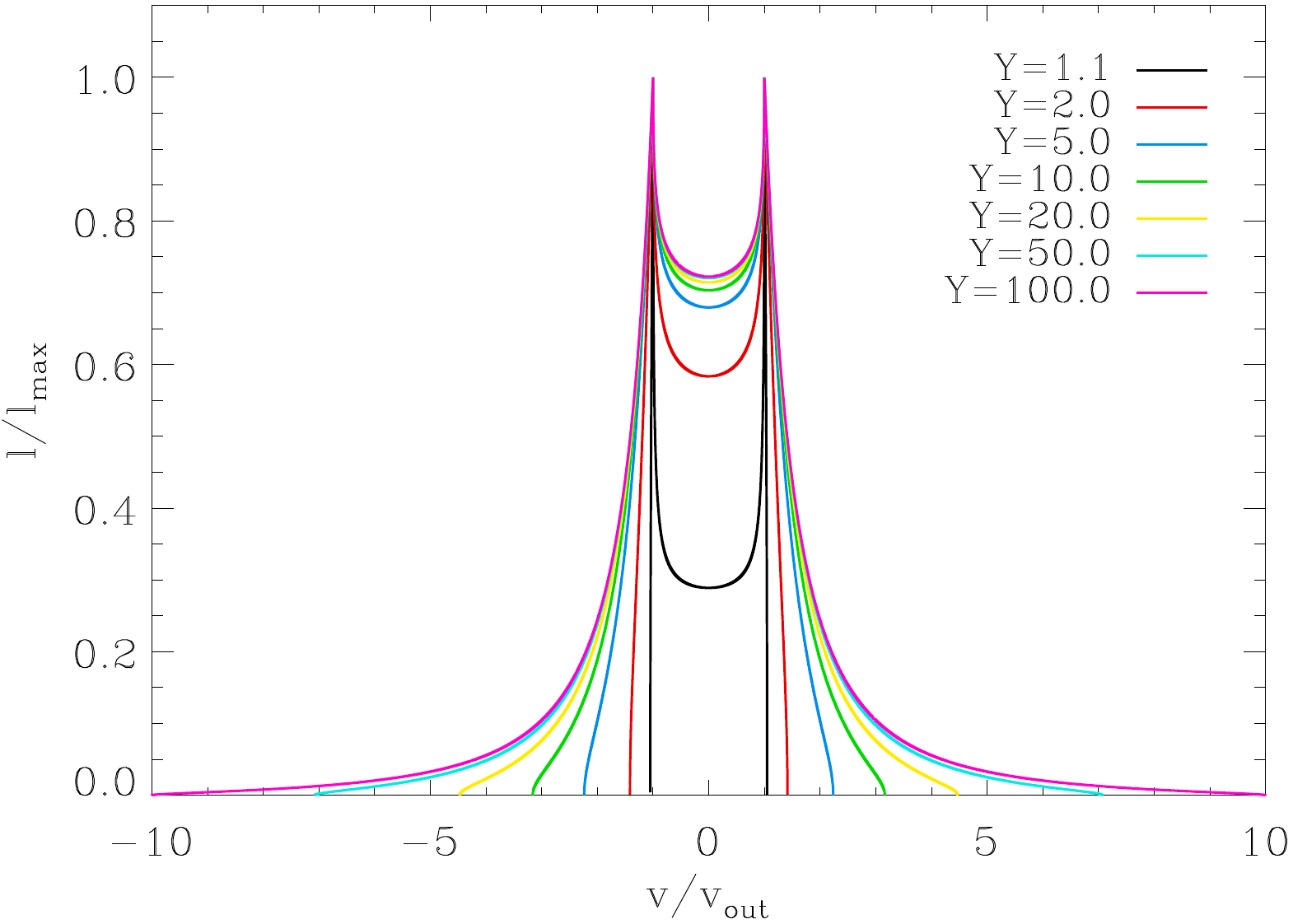}
\caption{Lengths of iso-velocity contours for a Keplerian disk as a function of
line-of-sight velocity for various values of the disk width $Y = \Ro/\Ri$.}
\label{fig:lengths}
\end{figure}

\begin{figure}
\centering
  \includegraphics[width=\hsize,clip=]{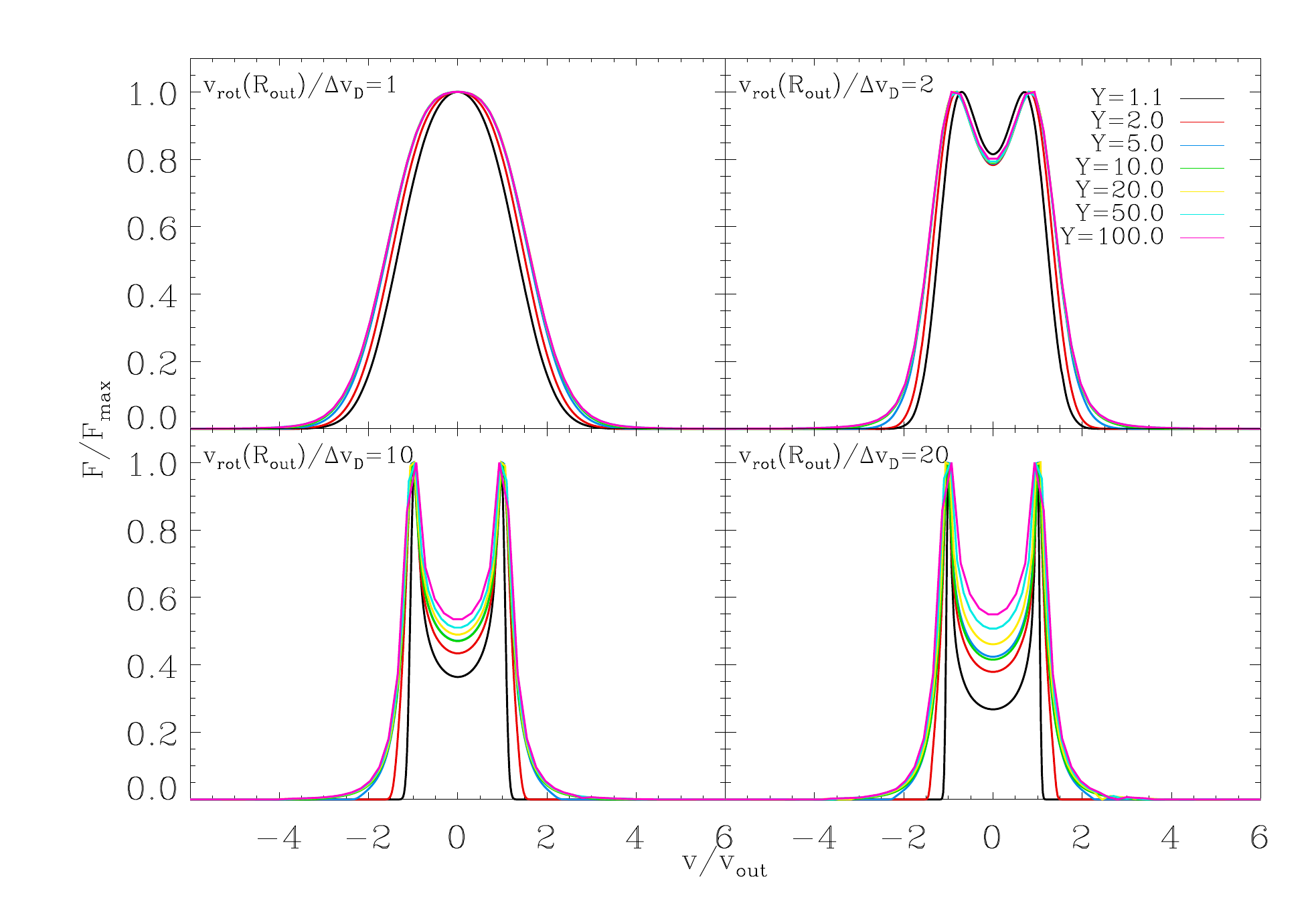}
\caption{Spectral shape of the emission from Keplerian disks with Gaussian line
profile $\Psi\sub{G} = e^{-x^2}$ at increasing rotation velocity. The different
curves in each panel cover a range of disk widths, as marked. These results
should be applicable to optically thin disks (see eq.\ \ref{eq:Gauss_range}). }
\label{fig:gaussian}
\end{figure}

The variation of contour length with los velocity captures the essence of
kinematic double-peaked profiles in Keplerian disks: \vo\ has the longest
contour and thus produces peak emission. Figure \ref{fig:gaussian} shows
profiles calculated from eq.\ \ref{eq:basic} with a Gaussian shape for the
intrinsic line profile $\Psi$ and increasing rotation velocity. The sequence is
best understood by considering separately the emission from the approaching
(left; fig.\ \ref{fig:Contours}) and receding (right) halves of the rotating
disk. With increasing rotation speed, the emission from each half is sliding
outward in \v, away from \v\ = 0. As long as $\vo \le \Dv$, the emissions from
both halves largely overlap and there are no double peaks. Increasing further
the rotation velocity, kinematic double peaks emerge, centered on $\v =
\pm\vo$, and the plots begin to resemble the behavior of the length contours
(fig.\ \ref{fig:lengths}). In particular, the profiles for $\vo = 20\Dv$ are
quite similar to the length curves, and the depth ratio \Imin/\Imax\ for the
central dip of a narrow disk (ring) with $Y = 1.1$ is nearly the same as the
corresponding length ratio. As the disk width increases, both the contour
lengths and the emission profiles show quick transitions to limit ratios,
albeit at different values. From analytic expressions developed in Appendix
\ref{sec_ap:disk} we find that
\eq{\label{eq:contrast}
    {\Imin\over\Imax} \simeq \cases{ \sqrt{Y - 1}           & $Y < 2$ \cr \cr
                                         4/3\pi             & $Y > 2$}
}
Only narrow disks ($\Ro < 2\Ri$) can produce kinematic double peaks with a
central dip deeper than 0.4.

The Gaussian-profile results are applicable to optically thin lines and can be
used to assess the disk parameters that will produce unequivocal kinematic
double peaks. Since $\vo = \vrot(\Ro)\sin i$ (eq.\ \ref{eq:vrot}), figure
\ref{fig:gaussian} shows that double peaks require not only $\vrot(\Ro) > \Dv$
but also $\sin i > \Dv/\vrot(\Ro)$. And since the los optical depth is $\T(i) =
\T/\cos i$, the requirement of optically thin emission implies $\cos i > \T$.
Taken together, the parameter region
\eq{\label{eq:Gauss_range}
   \T < 1, \quad  {\vrot(\Ro)\over\Dv} > 1, \quad
   \arcsin{\Dv\over\vrot(\Ro)} < i < \arccos \T
}
will always produce kinematic double peak profiles.


\begin{figure}
\centering
  \includegraphics[width=\hsize,clip=]{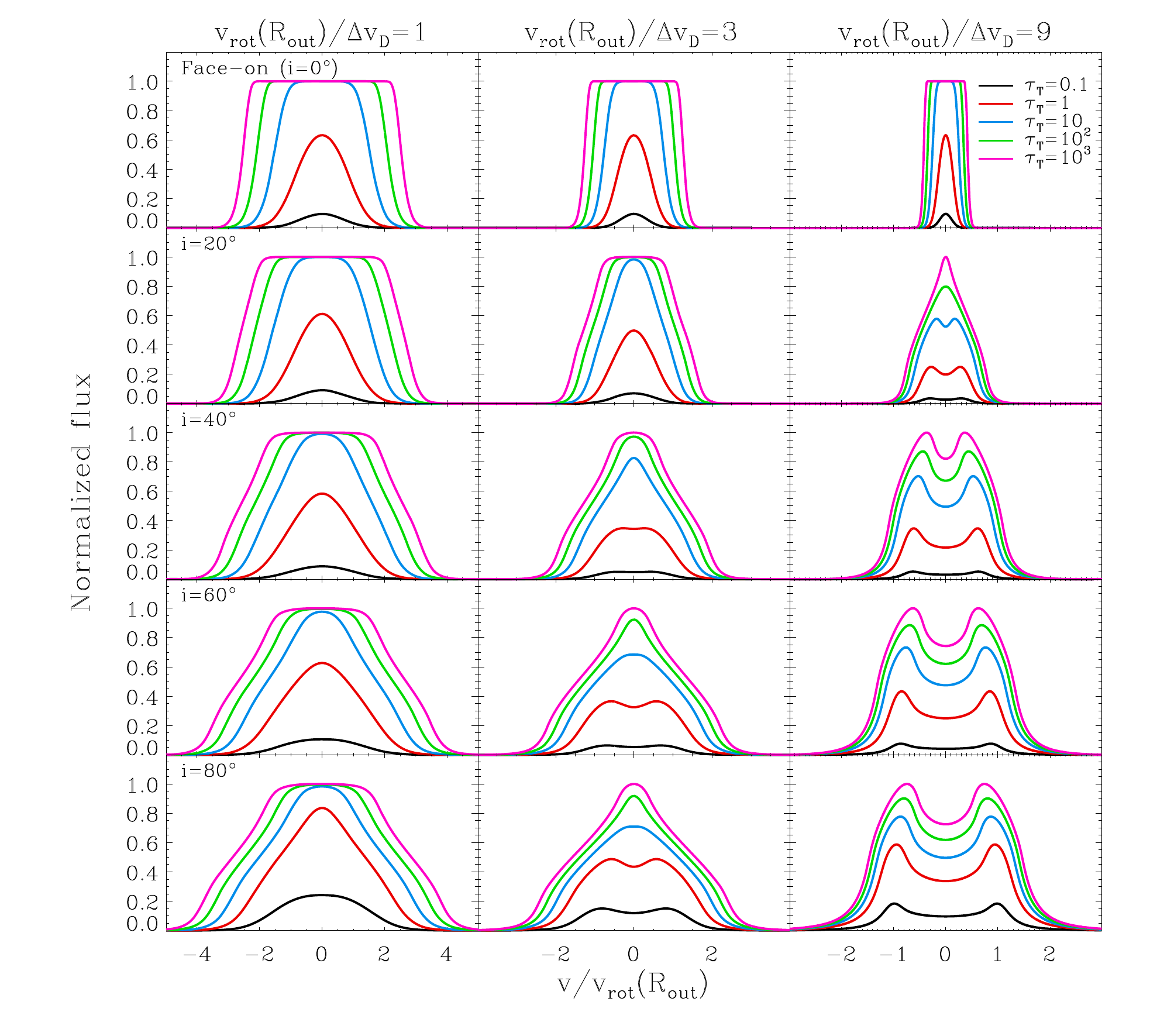}
\caption{Spectral shape of the emission from Keplerian disks with the Horne \&
Marsh profile $\Psi\sub{HM} = 1 - e^{-\tau(x)}$. These results should be
applicable to sources with $n \ge \nc$ ($\epsilon \ge 0.5$; see
\S\ref{sec:Thick_disk}). The disk radial thickness is \Ro/\Ri\ = 10, the
rotation velocity ranges from \vrot(\Ro) on the outer edge to $\vrot(\Ri) =
\sqrt{10}\,\vrot(\Ro)$ on the inner one. Rows correspond to various
inclinations, as marked, columns show a progression of rotation velocities; the
apparent profile narrowing between columns arises from the scale change of the
velocity axis. In each panel, line colors correspond to different values of \T,
the vertical optical depth, as marked, and the emission is normalized to a
maximum of unity.
} \label{fig:HM_vout}
\end{figure}


\subsection{Optically Thick Disks}
\label{sec:Thick_disk}

Optical depth effects fall into two regimes according to the relation between
the density and the transition critical density. When $n \ge \nc$ ($\epsilon
\ge 0.5$; eq.\ \ref{eq:epsilon}) the spatial variation of the source function
is sufficiently small that the dip-to-peak contrast of the non-kinematic double
peaks is diminished. Then the effect of increasing optical depth is mostly to
flatten the top of the intrinsic emission profile and increase its width by a
factor of $\about \sqrt{\ln\T}$ (see \S\ref{sec:Slab}). In this regime the
profile $\Psi\sub{HM} = 1 - e^{-\tau(x)}$ used in the \cite{Horne86}
calculations is a reasonable approximation for the exact solution of the line
transfer problem. Figure \ref{fig:HM_vout} presents calculations with this
profile, demonstrating the impact of line broadening with increasing optical
thickness of the Keplerian disk. Because of the dependence of los optical depth
on $i$, the viewing angle now affects not only the los velocity but also the
intrinsic profile shape. Whereas the viewing angle effect was fully absorbed
into \vo\ in the two previous figures, now it must be shown explicitly. The
columns in figure \ref{fig:HM_vout} show the effect of increasing rotation
velocity while the rows show the effect of increasing viewing angle. The top
row shows disks viewed face-on, demonstrating the profile broadening with
vertical optical depth\footnote{The row's three panels are identical; the
apparent narrowing with rotation velocity simply reflects the scaling of the
$x$-axis with \vrot(\Ro).}. In addition to its impact on the profile shape, the
optical depth affects also the intensity scale---emission from optically thin
disks is reduced by a factor of $\T/\!\cos i$ from that of optically thick
disks with the same source function.

\begin{figure}
 \centering
\includegraphics[width=\hsize]{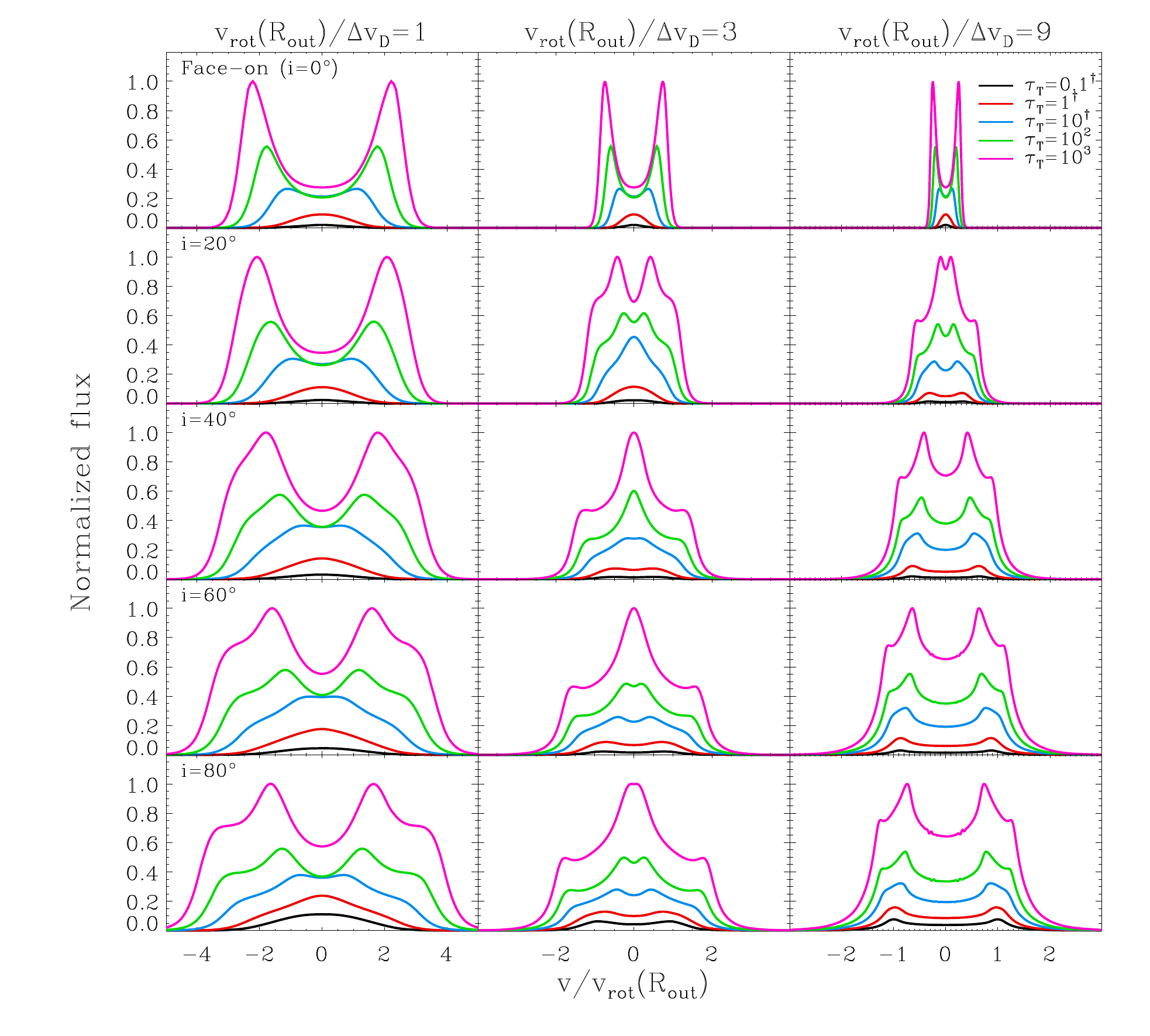}
\caption{Frequency profiles, normalized to unity peak in each panel, of line
emission from uniform, flat Keplerian disks. The intrinsic emission profile
$\Psi(x)$ is obtained from the radiative transfer solution for a slab with
photon destruction probability $\epsilon = \E{-2}$ and optical depth as marked.
All other disk properties are the same as in figure \ref{fig:HM_vout}. For
clarity, the results for \T\ = 0.1, 1 and 10, marked by $\dag$, were multiplied
by 15, 5 and 2, respectively.}
 \label{fig:flat.001}
\end{figure}

When $n < \nc$ ($\epsilon < 0.5$), optical depth effects introduce
non-kinematic double peaks in addition to broadening the emission spectral
range (\S\ref{sec:Slab}). This region requires full solution of the radiative
transfer problem, and we utilize the emission profiles $\Psi(x)$ from the slab
solution (see fig. \ref{fig:slab}) in the fundamental relation for emission
from rotating disks (eq.\ \ref{eq:basic}). Figure \ref{fig:flat.001} shows the
results for uniform disks with $\epsilon = \E{-2}$ and various optical depths,
rotating at increasing velocities and viewed from different directions; these
results are representative of the $\epsilon < 0.5$ domain. As in figure
\ref{fig:HM_vout}, the intensity is lower in the optically thin region because
of the smaller emitting column. Additionally, the intensity at low optical
depths is now further reduced because of the decrease in line excitation
temperature, reflecting the sub-thermal level populations. The face-on disks
shown in the top row replicate the behavior seen in fig.\ \ref{fig:slab}:
Gaussian profiles when $\T \le 1$, non-kinematic double peaks for the optically
thick disks.

\begin{figure}
 \centering
\includegraphics[width=0.8\hsize]{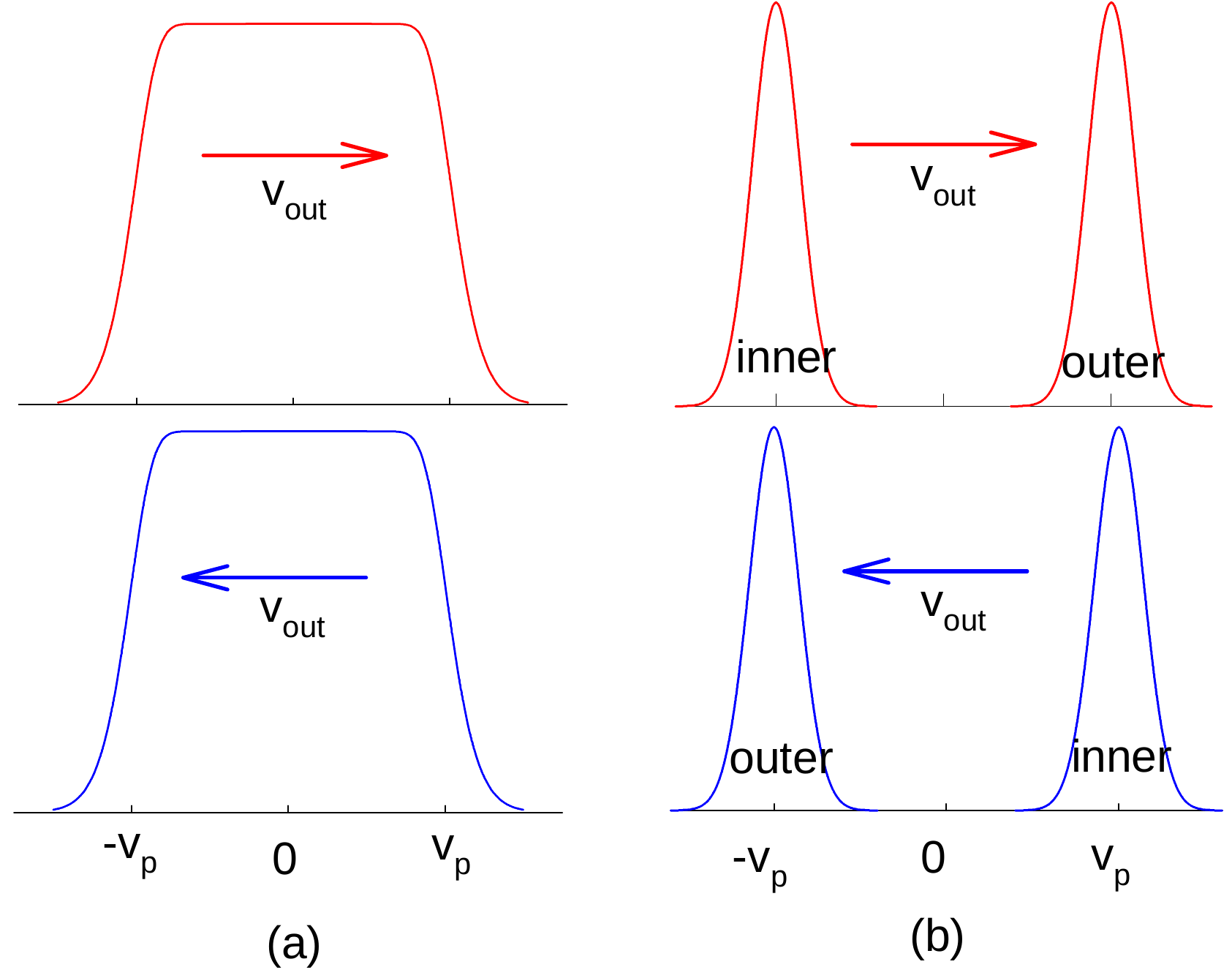}
\caption{Disk rotation and optically thick line emission. Shown are schematic
illustrations of the stationary-disk profiles when ({\it a}) $n > \nc$ and
({\it b}) $n < \nc$. In both cases, rotation shifts to the right (left) the
emission from the receding (approaching) half of the disk by the amount \vo. In
({\it b}), the leading non-kinematic peaks are labeled ``outer'', the trailing
ones ``inner''. The outer peaks are always moving away from each other, the
inner ones are first approaching and then receding as \vo\ increases.
} \label{fig:sketch}
\end{figure}

The effects of rotation on observed profiles at slanted viewing are broadly
similar in both figures \ref{fig:HM_vout} and \ref{fig:flat.001} and can be
readily understood with the aid of the illustrations in figure \ref{fig:sketch}
that sketch schematically the disk intrinsic profile in each case. Introduce
$\vp = \Dv\sqrt{\ln(\T/\cos i)}$. Then the $n > \nc$ profile can be described
with a flat top for $|\v| < \vp$ (figure \ref{fig:sketch}a) while the intrinsic
$n < \nc$ profile can be taken as just two peaks with centers at $\v = \pm\vp$
and width \about\ \Dv\ (fig. \ref{fig:sketch}b). The top and bottom of each
panel illustrate the emission from each half of a stationary disk. When the
disk is set in rotation, the profiles of the two halves slide against each
other in opposite directions away from the center \v\ = 0; the emission from
the approaching half (blue profile) is shifted by \vo\ (eq.\ \ref{eq:vrot}) to
the left, from the receding half (red) to the right by the same amount. Full
kinematic separation of the emission from the two halves of the disk occurs
only when $\vo > \vp + \Dv$ and the trailing edge of each profile crosses the
center. Thus the criterion for emergence of kinematic double structures is now
\eq{\label{eq:HM_range}
    {\vrot(\Ro)\over\Dv}\sin i > \sqrt{\ln(\T/\cos i)} + 1
}
This replaces eq.\ \ref{eq:Gauss_range} as the condition for kinematic double
peaks when the disk is optically thick rather than thin. At any given \T\ ($>
1$), this condition can be met only for sufficiently large viewing angles and
rotation speeds. For $\T > 10$ this criterion is met in the right column panels
of figures \ref{fig:HM_vout} and \ref{fig:flat.001}, where \vrot(\Ro) = 9\Dv.
At lower rotation velocities the blue and red profiles partially overlap,
resulting in the central peak evident in the mid-columns of figures
\ref{fig:HM_vout} and \ref{fig:flat.001} in the $i \ge 40\deg$ panels. The
central peak is produced for rotation velocities in the range $\vp - \Dv < \vo
< \vp + \Dv$, namely
\eq{\label{eq:central_peak}
        \sqrt{\ln(\T/\cos i)} - 1 < {\vrot(\Ro)\over\Dv}\sin i
                    < \sqrt{\ln(\T/\cos i)} + 1
}
and it covers the spectral overlap region $|\v| < \vo -\vp + \Dv$.

Because of its greater intrinsic complexity, the non-kinematic double-peaked
profile of $\epsilon < 0.5$ sources produces some additional structure in
rotating disks. In the schematic sketch of figure \ref{fig:sketch}, the leading
peak of the emission from each half of the disk is labeled ``outer'', the
trailing one ``inner''. Rotation shifts the outer peaks away from the center
and their separation increases monotonically with the rotation speed. These
peaks are discernible in the $i > 20\deg$ panels of figure \ref{fig:flat.001}
as the outermost bumps at $\v = \pm(\vo + \vp)$ in the profiles for all $\T \ge
10$; they are always the lowest structures, reflecting emission from only half
of the disk. The inner peaks first approach each other, then overlap to produce
the central peak (eq.\ \ref{eq:central_peak}) and finally move away from each
other once their trailing edges cross the center (eq.\ \ref{eq:HM_range}). This
final regime produces the quadruple-peak profiles evident in the right-column
panels for $\T > 10$ and $i \ge 20\deg$. Here the intrinsic non-kinematic
double-peaked profiles are superposed on the underlying kinematic double-peaks
with their centers shifted to $\pm\vo$. The kinematic separation produced by
rotation is centered on the dip between each pair of non-kinematic peaks, and
not on the prominent inner peaks; the markers of the rotation velocity are
these dips. Quad-peak profiles are seen also in the left column of figure
\ref{fig:flat.001} in the $i \ge 60\deg$ panels for $\T \ge 100$. These
quad-peaks occur at low velocities, $\vo < \vp - \Dv$, when the leading edge of
each inner peak has not yet reached the center. The two doublets making up the
quad are centered on $\pm\vp$, comprised each of the outer peak of emission
from one half of the disk and the inner peak from the other half. The
separation between the peaks of the two members of each doublet is 2\vo\ and
the separation between their prominent, inner peaks is $2(\vp - \vo)$. The
inner pair of peaks is higher than the outer one because they get contributions
from the two underlying non-kinematic peaks.

\section{Discussion}
\label{sec:discussion}

Double-peaked line profiles can arise from kinematic motions such as disk
rotation, collapse or bipolar outflow. In particular, rotating disks produce
such profile shapes for rotation velocities that obey the relations in eq.\
\ref{eq:Gauss_range} when $\T < 1$ or eq.\ \ref{eq:HM_range} when $\T > 1$.
However, double-peaked profiles arise also in the emission from stationary
optically thick sources when the density is below the critical density. There
is no obvious way to choose between these two possibilities from the profile
shape alone since non-kinematic double peak profiles of face-on disks with $\T
\ge 10$ (fig.\ \ref{fig:flat.001}) are quite similar to their kinematic
counterparts in optically thin slanted disks (fig.\ \ref{fig:gaussian}).
Because there is no clear-cut distinction between the two cases, a
double-peaked shape must be supplemented by additional information before it
can be taken as the signature of a rotating disk. Such independent input
generally involves the line optical depth. When a transition is known to be
optically thin in a given source, double-peaked emission can be reliably
attributed to kinematic effects and eq.\ \ref{eq:Gauss_range} then constrains
the source parameters. The depth of the central dip is limited in that case to
$\Imin/\Imax \ga 0.4$ in all but narrow disks (eq.\ \ref{eq:contrast}). A
deeper central dip implies a narrow, ring-like emission region with $\Ro <
2\Ri$.

When the possibility exists that a line is optically thick the situation is
more involved because double-peaked profiles can then arise from either
rotation or radiative transfer or both. If it is known additionally that $n \ge
\nc$ then double-peaks convincingly imply a kinematic origin because the
intrinsic profile is nearly flat top (fig.\ \ref{fig:HM_vout}). However, when
there is reason to believe that $n$ could be less than \nc\ then disk rotation
can be deduced only if it can be shown that the peak separation exceeds
$\Dv\sqrt{\ln(\T/\cos i)}$ for all reasonable estimates of \Dv\ and \T. While
this is not always possible, the triple and quadruple profiles seen in figure
\ref{fig:flat.001} can be used instead to identify optically thick rotating
disks. Indeed, in certain circumstances these complex profile shapes could be
more reliable indicators of rotation because they cannot be produced by
radiative transfer alone. The emergence of these profiles is a robust result
(see \S\ref{sec:Thick_disk}, especially fig.\ \ref{fig:sketch}) that should not
depend on the simplifying assumptions in our numerical calculations.

To demonstrate the formation of double-peaked profiles in the absence of any
large scale motions we presented here the simplest line transfer problems with
the minimal number of free parameters. In particular, we considered a single
two-level transition with line excitation purely by collisions in an
environment with constant temperature and density. However, the results remain
applicable for a much larger set of conditions because the central dip arises
from the decrease of the line excitation temperature away from the center
toward the surface of the source. This decrease arises purely from radiative
transfer effects in our simplified models, but it is expected in most
environments with variable physical conditions because temperature and density
generally decrease outward, amplifying the effect. A line excitation
temperature that instead increases toward the surface is possible only under
rather special circumstances, such as temperature inversion for example, and
our results would not be applicable in such cases. Similarly, our calculations
employed only the Gaussian line profile (implicitly assuming complete
redistribution) but the emergence of the central dip depends on the fact that
the line profile decreases toward the wings, not on the specific shape of this
decrease. The dip occurs because the emission at the line core follows the
spectral variation of $S[1/\Phi(x)]$ (see \S\ref{sec:Slab} and eq.\
\ref{eq:I(x)}), thus the Lorentzian profile would produce the dip just the
same, albeit with a somewhat different shape. Finally, the disk analysis
presented here was performed for simplicity in the thin disk approximation.
Taking account of the vertical variation of physical parameters requires a more
elaborate calculation that adds more free parameters \citep[e.g.][]{Andres07}
but is not expected to affect significantly the profile shapes found here. Our
results should remain valid as long as the viewing angle is such that the disk
edge contribution remains negligible and the absorption of the continuum does
not play an important role (as, for example, in the case of water lines:
\citealt{Cernicharo09, Ceccarelli10}). The calculations were conducted for
uniform disks, with $R$-independent line-center intensity $I_0(R)$. They are
applicable to other situations because a disk whose $I_0(R)$ declines steeply
with $R$ can be approximated with a uniform disk whose boundaries are limited
to the peak emission region. This explains the \cite{Horne86} modeling results
for the eclipsing dwarf nova Z Cha. The double peaked profile in this source
has $\Imin/\Imax = 0.4$, roughly the deepest dip possible for optically thin
Keplerian disks with $\Ro \ge 2\Ri$. While the \citeauthor{Horne86} model disk
had a much larger \Ro\ \about\ 10\Ri, it also had $I_0(R) \propto 1/R^2$. This
steep intensity decline ensures an effectively ring-like emission region that
enabled reproduction of the observed contrast.

Our results show that detection claims of disks based on double-peaked profiles
of molecular lines must be always scrutinized carefully for optical depth
effects. Double-peaked profiles of presumably optically thin lines (e.g. from
rare isotopes) can be used reliably for identifying candidate disks around YSO
whose average sizes are too small to be resolved even with ALMA. The case of
AGN, where double peaked H$\alpha$ and Mg\,II $\lambda$2798 profiles were
interpreted as rotating disk emission \citep{Eracleous09}, is more complicated.
Because their excitations are driven by photoionization and not collisions, our
model calculations are not directly applicable to these lines. Still, their
large optical depths, \about\ \E5--\E6\ (H.\ Netzer, private communication),
imply that the source function is unlikely to be constant in either case,
triggering the non-kinematic double peaked shapes seen in \S2. An actual
estimate of this effect will require detailed photoionization model
calculations coupled with exact radiative transfer calculations for these lines
instead of the common escape probability approach.

\appendix

\section{Slab Analytic Approximations}
\label{sec_ap:slab}

Neglecting radial coupling, the radiative transfer problem of a face-on flat
disk is identical to that of a slab, as noted above (\S\ref{sec:Disk}). Here we
present some approximate analytic expressions that reproduce adequately the
solution of the slab problem.

\begin{figure}
\centering
\includegraphics[width=\hsize]{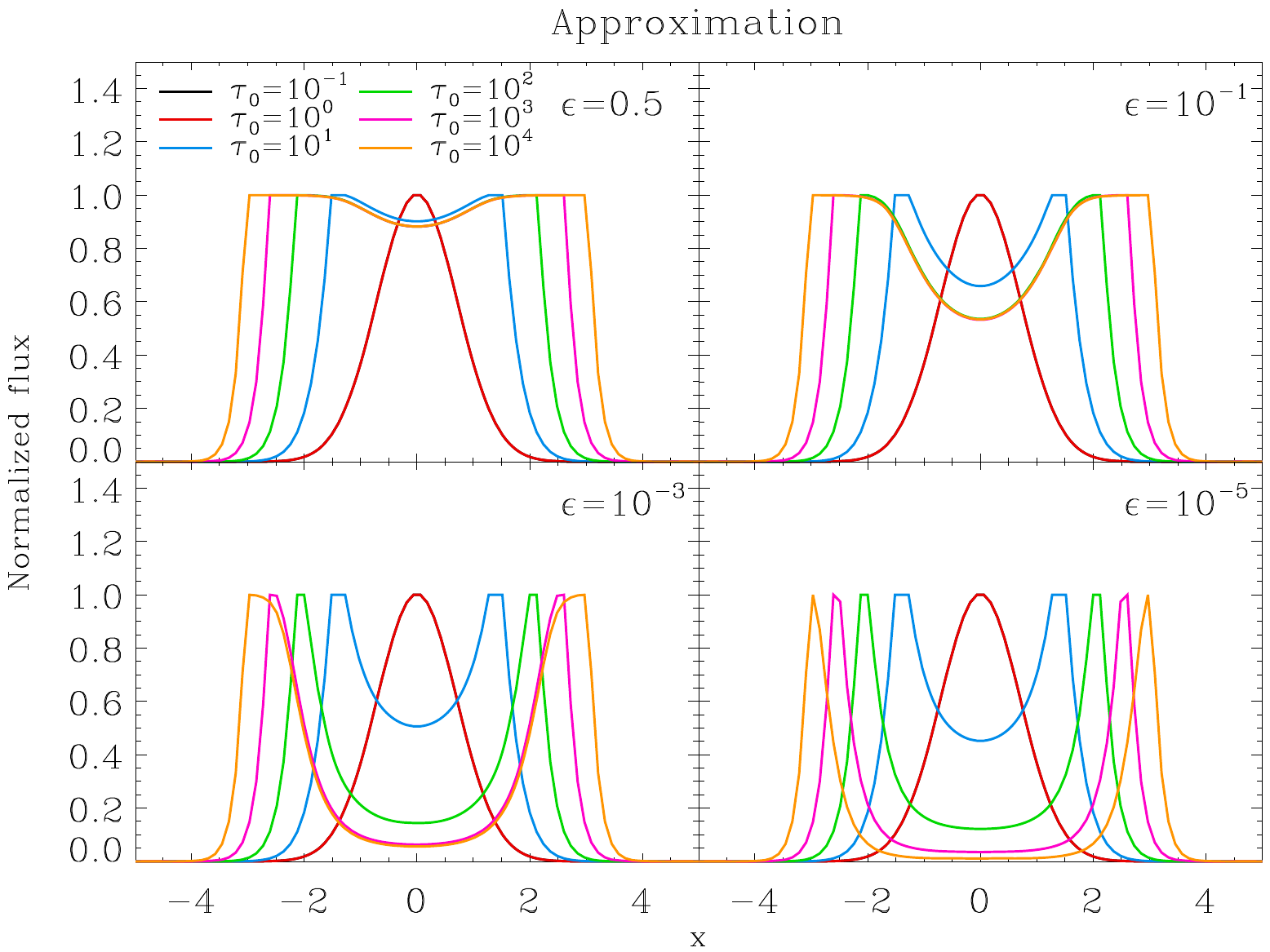}
\caption{Line radiation emergent from slabs with the parameters of the models
displayed in figure \ref{fig:slab}. These profiles were calculated analytically
by combining the approximate expressions in equations \ref{eq:I(x)} through
\ref{eq:S_fits}. \label{fig:profiles}}
\end{figure}

The two-level problem in a flat uniform slab is fully defined by the photon
destruction probability $\epsilon$ and the total optical depth \T\ across the
slab. Positions in the slab are characterized by the vertical distance $\tau$
from one face, $0 \le \tau \le \T$. The formal solution for the intensity
emerging perpendicular to the slab face is
\eq{
   I(x) = \Phi(x)\int_0^{\tau_{\rm T}} S(\tau)
             e^{-(\tau_{\rm T} - \tau)\Phi(x)}d\tau
}
Whenever $\T\Phi(x) < 1$,
\eq{
            I(x) \simeq \Phi(x)\int_0^{\tau_{\rm T}} S(\tau)d\tau
               = \Sbar \T\Phi(x)
}
where \Sbar\ is the value of the source function at some (a-priori unknown)
point inside the slab. We approximate the emission in the optically thin wings
($\T\Phi(x) < 1, \ |x| > \xw$; see eq.\ \ref{eq:x1}) with this expression. In
the optically thick core ($\T\Phi(x) > 2, \ |x| < \xc$), according to the
Eddington-Barbier relation the radiation emerges from an optical depth of
\about\ 1 from either surface, therefore the intensity is approximately $S[\tau
= 1/\Phi(x)]$. The frequency range $\xc \le |x| \le \xw$, where \hbox{$2 \ge
\T\Phi(x) \ge 1$}, does not lend itself naturally to any physical approximation
since the optical path is thick to one surface and thin to the other (except
for the midplane where both paths are optically thin). However, this is a small
spectral region whose relative thickness ($\xw - \xc$)/\xc\ decreases with
overall optical depth (eq.\ \ref{eq:x1}), therefore we approximate the
intensity there with a constant. Joining smoothly the core and wing
approximations with this constant yields $\Sbar = S(\T/2)$ --- the unknown
\Sbar\ is the source function on the slab midplane.  Thus our approximation for
the emergent radiation is
\eq{\label{eq:I(x)}
   I(x) \simeq \left\{
        \begin{array}{lll}
            S[1/\Phi(x)]         & \qquad          & |x|  <  \xc  \\
            S(\T/2)              & \qquad  \xc \le & |x| \le \xw  \\
            S(\T/2)\,\T\Phi(x)   & \qquad          & |x| > \xw
        \end{array}
               \right.
}
Completion of the intensity expression requires an approximation for the
$\tau$-variation of $S$ inside the slab. Motivated by Eq.\ 2.18 of
\cite{Avrett65}, we fitted the source function throughout the slab with the
following empirical expression containing the free parameters $a$, $b$ and
$\alpha$:
\eq{\label{eq:fit_source_slab}
    S(\tau; \epsilon, \T) = a + b k_1(\epsilon^\alpha \tau),
}
where
\[
    k_1(\tau) = 1 - \frac{1}{2}
    \left[ K_2(\tau) + K_2(\T-\tau) \right]
\]
and
\[
    K_2(\tau) = {1\over\sqrt{\pi}}\int_{-\infty}^\infty
                \Phi(x) E_2 \left[ \tau \Phi(x) \right] \mathrm{d}x,
\]
with $E_2$ the 2nd exponential integral. We find that $\alpha = 0.7$ yields
satisfactory agreement with the exact numerical solutions for all values of
$\epsilon$ and \T. The constants $a$ and $b$ are related to the values of the
source function on the slab surfaces and its midplane via
\begin{eqnarray}
 a &=& S(0) - b\,k_1(0)                                         \\ \nonumber
 b &=& \frac{S(\T/2) - S(0)}{k_1(\epsilon^\alpha\T/2) - k_1(0)}
\end{eqnarray}
The values of $S(0)$ and $S(\T/2)$ can be approximated utilizing some general
properties of the slab radiative transfer problem derived in \cite{Avrett65}.
They show that the source function on the surface of a slab with an overall
optical depth \T\ can be related to that at depth \T\ into a semi-infinite
atmosphere. For the latter problem there exist some general theorems in the
limits of both large and small $\epsilon$. Utilizing those, we devised the
following simple relations:
\begin{eqnarray}\label{eq:S_fits}
  S(\T/2) &=& \frac{\epsilon + \epsilon\T} {1 + \epsilon \T}\, B \\ \nonumber
  S(0)    &=& \sqrt{\epsilon} \frac{\sqrt{\epsilon} +
          \beta_1(\epsilon\T)^{p_1}+\beta_2(\epsilon \T)^{p_2}}
          {1 + \beta_1(\epsilon \T)^{p_1} + \beta_2(\epsilon \T)^{p_2}}\, B,
\end{eqnarray}
where $B$ is the Planck function of the (constant) slab temperature. We find
that $S(0)$ is reproduced satisfactorily with $\beta_1 = 1.41$, $\beta_2 =
1.23$, $p_1 = 0.6$ and $p_2 = 0.55$. This completes our analytic approximation
for $S(\tau)$ in all slabs. Inserting this expression in the approximation for
the emerging intensity (eq.\ \ref{eq:I(x)}) produces the line profiles shown in
figure \ref{fig:profiles}. These profiles are calculated for the same
parameters used in the exact model calculations presented in figure
\ref{fig:slab}. Comparison of the two verifies the adequacy of the analytic
approximation provided by the combination of equations \ref{eq:I(x)} through
\ref{eq:S_fits}.


The depth of the central dip in double-peaked profiles can be characterized by
the ratio of intensities at line center and at the peaks. From eq.\
\ref{eq:I(x)},
\eq{\label{eq:dip}
    {\Imin\over\Imax} = {I(x = 0)\over I(x = \xc)}
        \simeq {S(\tau = 1) \over S(\tau = \T/2)}
}
This expression provides an excellent approximation for the exact results shown
in figure \ref{fig:Dip} for the variation of the central dip with optical depth
for various values of $\epsilon$, but $S(\tau = 1)$ is not easily discernible
from the somewhat involved approximation for the source function. Neglecting
the variation of $S$ close to the slab surface and replacing $S(1)$ with
$S(0)$, which is given explicitly in eq.\ \ref{eq:S_fits}, yields the simpler,
though somewhat less accurate, result
\eq{\label{eq:dip2}
    {I(x = 0)\over I(x = \xc)}
        \approx {S(\tau = 0) \over S(\tau = \T/2)}
}
Combining this relation with eq.\ \ref{eq:S_fits} produces the approximation
given in eq.\ \ref{eq:contrast0}.

\section{Keplerian Contours}
\label{sec_ap:disk}

All points on the disk surface whose emission is centered on the same frequency
\v\ lie on the contour defined in eq.\ \ref{eq:contours}. The polar angle
varies on the contour between the inner and outer radii in the range
[\tin,~\tout], where
\eq{\label{eq:tin-tout}
   \tin  =         \arccos{\v\over\vi}, \qquad
   \tout =  \cases{\arccos{\v\over\vo} & $0     \le \v \le \vo$ \cr \cr
                                 0       & $\vo \le \v \le \vi $ \cr}
}
Straightforward integration produces the contour length
\eq{\label{eq:ell}
 \ell(\v) = \Ro\left({\vo\over\v}\right)^2
                 \left[f(\tin) - f(\tout) \right]
}
where
\[
   f(\theta) = \sin\theta\sqrt{1 + 3\sin\theta^2} +
  {1\over\sqrt{3}}\ln\left(\sqrt{3}\sin\theta + \sqrt{1 + 3\sin\theta^2}\right)
\]
This relation produces the profiles shown in figure \ref{fig:lengths}.
Comparison with fig.\ \ref{fig:gaussian} shows good agreement with the results
of full calculations with narrow-width Gaussian shape for the intrinsic line
emission. In particular, the depth \Imin/\Imax\ of the central dip is nearly
the same for the narrow disk (ring) with $Y = 1.1$ as the length ratio
\lmin/\lmax\ for the same $Y$. Therefore, we can use the contour length for an
estimate of the central dip depth contrast in narrow disks:
\eq{
    {\Imin\over\Imax}  \mathrel{{\mathop\simeq_{Y \to 1}}} {\lmin\over\lmax}
}
The longest contour, \v\ = \vo, has \tout = 0 and $\tin = \arccos Y^{-1/2}$ so
that
\eq{
   \lmax = \Ro f\left(\sqrt{\arccos Y^{-1/2}}\right)
     \mathrel{{\mathop \rightarrow_{Y \to 1}}}
     2\Ro \sqrt{Y - 1}
}
while the length of the shortest contour, \v\ = 0, is simply
\eq{
   \lmin = 2\Ro(Y - 1).
}
Combining the last three relations produces the narrow-disk result listed in
eq.\ \ref{eq:contrast}.

Although the length calculation reproduces reasonably well the profile shape
and $Y$-dependence also for wide disks ($Y \gg 1$; see figures
\ref{fig:lengths} and \ref{fig:gaussian}), it misses on the actual value of
\Imax/\Imin, the dip contrast; the length calculation produces accurately the
central contrast for narrow disks but not for the thicker ones because of a
fundamental issue with the approach to a $\delta$-function limit. The line
profile can be considered narrow when $\Dv \ll \v$, and this condition is never
applicable at line center, where \v\ = 0. Since the line wings of the closed
contours contribute more to the $\v > 0$ contours of wide disks, the length
calculation misses the proper result; instead, a calculation of the emitting
area at velocity \v\ is necessary. The area under the \v-contour in the 1st
quadrant is
\eq{
   A(\v) = \frac12 R_{\rm out}^2\left({\vo\over\v}\right)^4
           \left[g(\tin) - g(\tout)\right]
}
where
\[
   g(\theta) =   \frac38\theta
               + \frac38\sin\theta\cos\theta + \frac14\sin\theta\cos^4\theta
\]
At large $Y$ we have $\tin \simeq \pi/2$ for the \vo-contour (eq.\
\ref{eq:tin-tout}) so that $g(\tin) = {3\over16}\pi$ and the area under
contours close to the longest one is
\eq{
   \v \simeq \vo:\quad
        A(\v) \simeq {3\pi\over32}\, R_{\rm out}^2 \left({\vo\over\v}\right)^4
}
For a rectangular velocity profile with width $\Dv \ll \vo$, it is
straightforward to show from this expression that the area of peak emission in
the 1st quadrant is
\eq{
 A_{\rm peak} = A(\vo - \Dv) - A(\vo) = {3\pi\over8}\,R_{\rm out}^2 {\Dv\over\vo}
}
At line center (\v\ = 0) the emitting area in the first quadrant is between the
vertical axis and a roughly radial ray with \v\ = \Dv. This ray has inclination
\Dv/\vo\ to the $y$-axis so that the area of dip emission is
\eq{
   A_{\rm dip} = \frac12\, R_{\rm out}^2 {\Dv\over\vo}
}
Therefore, the limiting depth is
\eq{\label{eq:A-ratio}
    {\Imin\over\Imax}  = {A_{\rm dip}\over A_{\rm peak}}
          \mathrel{{\mathop\simeq_{Y \gg 1}}} {4\over3\pi}
}
This result, listed in eq.\ \ref{eq:contrast}, reproduces fairly accurately the
depth contrast of the narrowest Gaussian profile ($\Dv/\vo = 1/20$) in figure
\ref{fig:gaussian}.




\label{lastpage}

\end{document}